\documentclass{article}
\usepackage{spconf,amsmath,graphicx}
\usepackage{cite}
\usepackage{multicol,multirow}

\title{Conversational Speech Recognition by Learning \\Conversation-level Characteristics}
\name{Kun Wei$^{1,2*}$, Yike Zhang$^{2*}$, Sining Sun$^2$, Lei Xie$^{1\dag}$, Long Ma$^2$}
\address{$^1$Audio, Speech and Language Processing Group (ASLP@NPU), School of Computer Science,\\ Northwestern Polytechnical University, Xian, China\\
     $^2$Cloud Xiaowei, Tencent, Beijing, China
     \thanks{$^*$Equal contribution.
     }
     \thanks{$^{\dag}$Corresponding author.}
     }

\begin{document}
%
\maketitle
\begin{abstract}
Conversational automatic speech recognition (ASR) is a task to recognize conversational speech including multiple speakers. Unlike sentence-level ASR, conversational ASR can naturally take advantages from specific characteristics of conversation, such as role preference and topical coherence.
This paper proposes a conversational ASR model which explicitly learns conversation-level characteristics under the prevalent end-to-end neural framework. The highlights of the proposed model are twofold. 
First, a latent variational module (LVM) is attached to a conformer-based encoder-decoder ASR backbone to learn role preference and topical coherence. Second, a topic model is specifically adopted to bias the outputs of the decoder to words in the predicted topics.
Experiments on two Mandarin conversational ASR tasks show that the proposed model achieves a maximum 12\% relative character error rate (CER) reduction.
\end{abstract}
\section{Introduction}
\label{sec:intro}

A typical automatic speech recognition (ASR) system usually works at sentence-level. It is trained by sentence-level speech-text paired data and recognize speech at (short) utterance level. In contrast, conversational ASR has great potential to take advantages from specific characteristics of multi-speaker conversation.
Role preferences, such as style and emotion, will affect the characteristics of the current conversation\cite{liang2021modeling}.
Topical coherence, 
the tendency of words that are semantically related to one or more underlying topics to appear together in the conversation,
and other conversation-level phenomena have also received widespread attention~\cite{xiong2018session}. 
Previous works have explored long context language models~\cite{irie2019training}, 
longer input features~\cite{hori2020transformer, hori2021advanced},
context attention mechanisms~\cite{kim2019cross} and other methods~\cite{masumura2021hierarchical} to implicitly learn contextual information in conversions~\cite{kim2018dialog, masumura2021hierarchical}. 
They do not explicitly make use of the inherent characteristics of conversations, 
such as role preference, topical coherence, speaker turn-talking, etc.
However, 
learning conversational characteristics in explicit ways may further improve performance of conversational ASR.

In this paper, 
we propose a conversational ASR model,
which learns conversational characteristics through 
a Conditional Variational Auto-Encoder (CVAE)~\cite{sohn2015learning} 
and a topic model~\cite{bianchi-etal-2021-pre}. 
Inspired by~\cite{liang2021modeling}, 
we use 
a role-customized variational module and a topic-customized variational module to obtain the characterization of role preference and topical coherence respectively.
Additionally,
a Combined Topic Model (CombinedTM)~\cite{bianchi-etal-2021-pre}, 
which has contextualized document embeddings and stronger ability to express topical coherence, is used to rescore the top-k outputs of the ASR model.


We carry out experiments on Mandarin two-speaker telephony conversations. Specifically, results on two datasets HKUST~\cite{liu2006hkust} and DDT show that the proposed method achieves impressively a maximum 12\% relative character error rate (CER) reduction. Nevertheless, the proposed method can be applied to conversations involving more speakers, such as multi-party meetings, as well as other languages.

\section{Related Work}
End-to-end ASR models are becoming more and more popular due to their excellent performance and lower construction difficulty. As the most influential sequence-to-sequence model family adopting multi-head attention to learn global information of sequence, Transformer~\cite{vaswani2017attention} and its variants~\cite{gulati2020conformer, wang2021efficient}, have recently received more attention due to their superior performance on a wide range of tasks including ASR~\cite{wang2019learning, raganato2018analysis, dong2018speech, karita2019comparative, luo2021simplified,chiu2018state, chan2016listen, vaswani2017attention, guo2021recent}.


A common idea for applying transformer-based models to the long sequential tasks,
such as conversational ASR, 
is to model the long-context information. 
Recently, 
long-context end-to-end models that can learn information across sentence boundaries have draw much interest in the fields of 
long-sequence prediction~\cite{beltagy2020longformer, zhou2021informer},
machine translation~\cite{wang2017exploiting, liang2021modeling} 
and speech recognition~\cite{masumura2019large, kim2018dialog, masumura2021hierarchical}. 
In ~\cite{kim2019cross}, 
a cross-attention end-to-end speech recognizer is designed to solve the problem of speaker turn-talking. 
Meanwhile, a model with CVAE for conversational machine translation is proposed in~\cite{liang2021modeling}.


\section{The Proposed Method}
\begin{figure}[ht]

\centering
\includegraphics[scale=0.25]{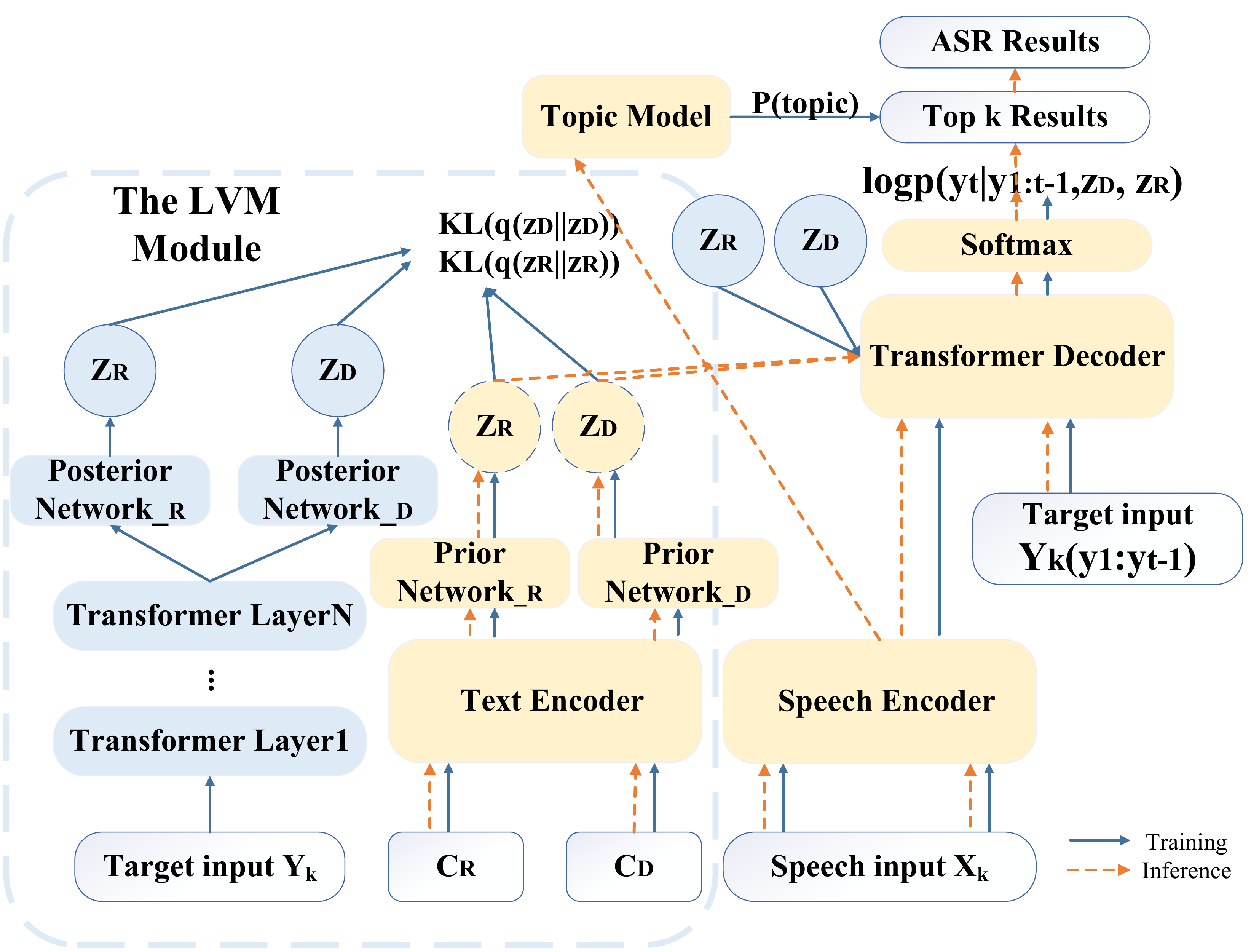}
\caption{
The overall framework of the proposed conversational ASR model.
Solid lines represent calculation paths at training, 
dashed lines represent calculation paths at inference,
subscript $D$ means topic, 
and subscript $R$ means role. }
\label{fig:CVAE}
\end{figure}

As shown in Figure \ref{fig:CVAE}, the proposed model consists of
a speech encoder, a latent variational module (LVM), a decoder and a rescoring module. 
First, 
speech input features $X_{k}$ and dialog embeddings ${C_\text{role}, C_\text{dia}}$ are sent into the speech encoder and the text encoder respectively. 
Then latent vectors $Z_\text{D}, Z_\text{R}$ derived from the variational modules are sent into the decoder to characterize topic and role information.
At training, 
latent vectors are derived from posterior networks in the variational modules. 
While at inference, 
latent vectors are derived from prior networks.
Finally, we use a topic model to rescore the output of the decoder. 
Each module in the proposed conversational ASR model will be elaborate as follows.

\subsection{Input Representation}
The input of our model consists of three parts at the $k$-th sentence in the conversation: 
the speech feature of current sentence $X_{k}$,
the target text $Y_{k}$ 
and the contextual input feature $\{C_\text{role}, C_\text{dia}\}$. 
Here, $C_\text{role}$ is transcripts of the current speaker, and $C_\text{dia}$ is all historical transcript in the current conversation. For example,
$C_{role}=(Y_{1}, Y_{3}, Y_{5},...,Y_{k-2})$ and $C_{dia}=(Y_{1}, Y_{2}, Y_{3},...,Y_{k-1})$. 
The text data is processed in a format in which each person speaks one sentence in turn, so role preference can also be expressed as $C_{role}=(Y_{2}, Y_{4}, Y_{6},...,Y_{k-1})$ for another speaker.
In order to distinguish different sentences, 
we add start symbol $<sos>$ and end symbol $<eos>$ at the beginning and ending of each sentence. 
Then, all text inputs will be represented as word embedding vectors.
\subsection{Speech Encoder}
Recently, Gulati et al. combined transformers and convolutional neural networks as Conformer~\cite{gulati2020conformer}
to simultaneously capture local and global contextual information in ASR tasks, leading to superior performance.
Here, we stack $N_\text{con}$ conformer blocks as our speech encoder.
Specifically, each conformer block includes 
a multi-head self-attention module (MHSA),
a convolution module (CONV) 
and a feed-forward module (FFN).
Assuming the input of the i-th block is $\textbf{h}_{i}$, 
operations in this block can be expressed as:
\begin{equation}
\label{Conformer}
\textbf{s}_{i} = \rm{MHSA}(\textbf{h}_{i}) + \textbf{h}_{i},
\end{equation}
\begin{equation}
\textbf{c}_{i} = \rm{CONV}(\textbf{s}_{i}),
\end{equation}
\begin{equation}
\textbf{h}_{i+1} = \rm{FFN}(\textbf{c}_{i}) + \textbf{c}_{i}.
\end{equation}

\subsection{Latent Variational Module}
The latent variational module consists of a text encoder and two specific VAEs: 
role VAE and topic VAE. 
Each VAE consists of multiple transformer layers, 
a posterior network
and a prior network, 
as shown in the left half of Figure \ref{fig:CVAE}. These two VAEs characterize  
role preference and topical coherence in conversations by learning role-specific latent vectors $Z_R$ and topic-specific latent vectors $Z_D$.

\noindent\textbf{Role VAE.} 
The structure of the text encoder (TEnc) is a standard transformer encoder~\cite{vaswani2017attention}. The intermediate representation vectors 
of role preference and topic consistency are generated by the same text encoder:
\begin{equation}
    \textbf{h}_\text{role}^\text{text} = \rm{TEnc}(Wd(C_{role})),
\end{equation}
\begin{equation}
    \textbf{h}_\text{dia}^\text{text} = \rm{TEnc}(Wd(C_{dia})),
\end{equation}
where Wd stands for word embedding operation. 
Then mean-pooling is applied to $\textbf{h}_\text{role}^\text{text}$ and $\textbf{h}_\text{dia}^\text{text}$ across time.
Here, 
$\textbf{h}_\text{dia} =\begin{matrix}\frac{1}{n}\sum_{i=1}^n {\textbf{h}_{\text{dia}, i}^\text{text}} \end{matrix}$, and $\textbf{h}_\text{role} =\begin{matrix}\frac{1}{n}\sum_{i=1}^n {\textbf{h}_{\text{role}, i}^\text{text}} \end{matrix}$, where $n$ is the length of the context sentence.
We use the historical text of the same speaker in previous turns of current dialog 
to generate the character variational representation $\textbf{z}_\text{role}$,
which follows an isotropic Gaussian distribution~\cite{wang2019t}, 
\begin{equation}
     p_{\theta}(\textbf{z}_\text{role}|C_\text{role})\sim N(\mu_\text{role}, \sigma^{2}_\text{role}\textbf{I}),
\end{equation}
where \textbf{I} denotes the identity matrix and
\begin{equation}
    \mu_\text{role} = \rm{MLP}^{role}_{\theta}(\textbf{h}_{role}),
\end{equation}
\begin{equation}
    \sigma_\text{role} = \rm{Softplus}(MLP_{\theta}^{role}(\textbf{h}_{role})).
\end{equation}
MLP is a linear layer and Softplus is the activate function.

At training, the posterior distribution conditions on sentences related to the current role. 
Through KL divergence, the prior network can learn a role-specific distribution by approximating the posterior network~\cite{sohn2015learning}. The variable distribution of the posterior network is extracted as
\begin{equation}
     q_{\phi}(\textbf{z}_\text{role}|C_\text{role}, Y_{k})\sim N(\mu^{\prime}_\text{role}, \sigma^{\prime2}_\text{role}\textbf{I}),
\end{equation}
where
\begin{equation}
    \mu^{\prime}_\text{role} = \rm{MLP}_{\phi}^{role}(\textbf{h}_{role}, \textbf{h}_{y}),
\end{equation}
\begin{equation}
    \sigma^{\prime}_\text{role} = \rm{Softplus}(\rm{MLP}_{\phi}^{role}(\textbf{h}_{role}, \textbf{h}_{y})),
\end{equation}
here $\textbf{h}_\text{y}$ is the vectors calculated by the transformer layers of posterior network.
The conditional probability $q_{\phi}$
is learned by the posterior network when training. 
However, we fit this distribution through a prior network at inference, so as to avoid the dependence on the recognition result of current sentence and the deviation of the recognition result from the real result. 

\noindent\textbf{Topic VAE.} In the topical coherence problem, we use a structure similar to the above to extract relevant representations of topical coherence $\textbf{z}_\text{dia}$.
\begin{equation}
     p_{\theta}(\textbf{z}_\text{dia}|C_\text{dia})\sim N(\mu_\text{dia}, \sigma^{2}_\text{dia}\textbf{I}),
\end{equation}
where \textbf{I} denotes the identity matrix and
\begin{equation}
    \mu_{dia} = \rm{MLP}^{dia}_{\theta}(\textbf{h}_{dia}),
\end{equation}
\begin{equation}
    \sigma_\text{dia} = \rm{Softplus}(\rm{MLP}_{\theta}^\text{dia}(\textbf{h}_\text{dia})),
\end{equation}
At training, the prior network learns the distribution of topical coherence information by approximating the posterior network. The variable distribution of the posterior network is extracted as
\begin{equation}
     q_{\phi}(\textbf{z}_\text{dia}|C_\text{dia}, Y_{k})\sim N(\mu^{\prime}_\text{dia}, \sigma^{\prime2}_\text{dia}\textbf{I}),
\end{equation}
where
\begin{equation}
    \mu^{\prime}_\text{dia} = \rm{MLP}_{\phi}^\text{dia}(\textbf{h}_\text{dia}, \textbf{h}_\text{y}),
\end{equation}
\begin{equation}
    \sigma^{\prime}_\text{dia} = \rm{Softplus}(\rm{MLP}_{\phi}^\text{dia}(\textbf{h}_\text{dia}, \textbf{h}_\text{y})).
\end{equation}

\subsection{Decoder}
We use an attention decoder in the proposed model. 
As shown in Figure \ref{fig:CVAE}, 
we get the latent variables$\{\textbf{z}_\text{role},\textbf{z}_\text{dia}\}$ either from the posterior networks or the prior networks. 
A transformer layer $\textbf{M}_\text{trans}$ is used to merge these intermediate vectors into the decoded states:
\begin{equation}
    \textbf{g}_{t} = \rm{Tanh}(\textbf{M}_\text{trans}(\textbf{h}_{s}, \textbf{z}_\text{role}, \textbf{z}_\text{dia})+\textbf{b}_\text{trans}),
\end{equation}
where $\textbf{h}_{s}$ is the hidden state of decoder. Then, we send $\textbf{g}_{t}$ to a softmax layer to get the probability of target chars.

\subsection{Training Objectives}
We adopt a two-stage training strategy. We first train a sentence-level speech recognition model with the cross-entropy objective:
\begin{equation}
    L_\text{ce}(\theta_{s}; X, Y) = -\sum_{t=1}^n \log{p_{\theta_{s}}(y_{t}|X, y_{1:t-1})}.
\end{equation}
Then, we finetune the model by minimizing $L_\text{ce}$ and the follow objective:
\begin{equation}
\begin{aligned}
& L_\text{vae}(\theta, \phi; C_\text{role}, C_\text{dia}, X, Y)  = \\                                       &+KL(q_{\phi}(\textbf{z}_\text{role}|C_\text{role},Y_{k})||p_{\theta}(\textbf{z}_\text{role}|C_\text{role})) \\
&+KL(q_{\phi}(\textbf{z}_\text{dia}|C_\text{dia},Y_{k})||p_{\theta}(\textbf{z}_\text{dia}|C_\text{dia}))\\
&-{E}_{q_{\theta}}[\log p_{\theta}(Y_{k}|\textbf{z}_\text{role}, \textbf{z}_\text{dia})].
\end{aligned}
\end{equation}

\begin{table*}
\centering
\vspace{-0.6cm}
\caption{CER comparation of different end-to-end models on two Mandarin datasets. 
ExtLM is an RNN LM, 
RoleVAE and TopicVAE are the proposed VAEs,
AttRes is attention rescoring, 
and TopicRes means the proposed topic rescoring.}
\label{tab:all_results}
\begin{tabular}{lcccccccc}
\hline
\textbf{Model} & \textbf{ExtLM} & \textbf{RoleVAE}& \textbf{TopicVAE} & \textbf{AttRes} & \textbf{TopicRes}& \textbf{HKUST} & \textbf{DDT/dev} & \textbf{DDT/test} \\ 
\hline
\hline
\multirow{8}{*}{\textbf{Conformer}}   
& - & - & - & - &-       & 20.25 & 23.43 & 22.71 \\ 
& Y & - & - & - &-       & 20.45 & 23.23 & 22.12 \\ 
& - & Y & - & - &-       & 19.94 & 22.70 & 22.13 \\ 
& - & - & Y & - &-       & 19.81 & 22.62 & 22.36 \\ 
& - & Y & Y & - &-       & 19.46 & 22.35 & 22.06 \\ 
& - & Y & Y & Y &-       & 19.32 & 20.35 & 20.13 \\ 
& - & Y & Y & Y &Y       & \textbf{19.19}        & \textbf{20.02}          & \textbf{19.96}           \\ 
& Y & Y & Y & Y &Y       & 19.31 & 20.12 & 21.05 \\ \hline

\textbf{H-Transformer} & -& - & - & - &-                      & 20.14 & 23.01 & 22.53 \\ \hline
\vspace{-0.8cm}
\end{tabular}
\end{table*}

\subsection{Topic Model Rescoring}
We rescore the output of the ASR model in the process of attention rescoring~\cite{yao2021wenet}.
Specifically, 
we classify conversations in the training set into $m$ topics by the topic model CombinedTM.
Each topic contains $j$ words like $(v_{1}^{1},v_{2}^{1},...,v_{j}^{1},...,v_{j}^{m})$. Keywords in all topics do not overlap each other. For the top-n sentences $(S_{1},...,S_{n})$ generated by the speech recognition model, we send them to the topic model trained by the transcripts of speech dataset, and get the probability vectors of the sentence attribution to each topic$(\textbf{d}_{1}, \textbf{d}_{2}, ..., \textbf{d}_{m})$, each vector has $m$ dimensions. 
For the $t$-th word $w_{n}^{t}$ in $S_{n}$, if $w_{n}^{t}$ in the keywords of $b$-th topic we generated, the score of word $w_{n}^{t}$ $s_\text{old}(w)$ is recalculated as
\begin{equation}
s_\text{new}(w) = s_\text{old}(w) \times ( 1 + d_{n,b}).
\end{equation}
At attention rescoring, 
we add $s_\text{new}(w)$ to the attention:
\begin{equation}
s_\text{final}(w) = s_\text{attn}(w) + s_\text{new}(w), 
\end{equation}
where $s_\text{attn}(w)$ is the score calculated by the rescoring decoder.

Then we use the new score to reorder the output sentences, 
calculate the final sentence score with $s_\text{sen} = \begin{matrix}\sum_{i=1}^t {s_\text{final}({w}_{i}})\end{matrix}$.
The sentence with the highest score is considered as the final recognition result.
\section{Experiments}
\label{sec:typestyle}
\subsection{Dataset}
We conduct our experiments on two Mandarin conversation datasets -- HKUST~\cite{liu2006hkust} and DATATANG (DDT).
The HKUST dataset contains 200 hours of speech data. 
The dev set is used to verify the recognition results. The DDT dataset contains 350 hours of speech data. The dev and test sets are used to verify the recognition results.

We obtain the topic boundary information and speaker information of each round of dialogues through the additional tags of the data to distinguish speakers and judge the conversion of the topic. For each corpus, the detail configurations of our acoustic features and Conformer model are same as the ESPnet Conformer recipes~\cite{watanabe2018espnet} $(\text{Enc} = 12, \text{Dec} = 6, d^\text{ff} = 2048, \text{H} = 4, d^\text{att} = 256)$.
We use 3653 and 3126 characters as output units for HKUST and DDT respectively.
\subsection{Implementation Details}
We train our models with
ESPnet~\cite{watanabe2018espnet}. Speed perturbation at ratio 0.9, 1.0, 1.1 with SpecAugment~\cite{park2019specaugment} is used for data augmentation. 
The baseline results are trained on independent sentence level, without speaker and  context information.

In our experiment, we use a 2-layer text encoder 
and a 2-layer transformer as the feature extractor of the VAEs as shown in the left of Figure \ref{fig:CVAE}.
Latent variables with 100 dimensions are used to represent speaker information and topic information respectively. 
In the rescoring experiments, 
we divide the conversations in HKUST into 50 topics.
For DDT, 
we only divide the conversations into 3 topics as the topics are highly coterminous. 
In addition, a session-level RNNLM~\cite{xiong2018session} based on the training set text is applied.

We reproduce a comparative model, Hierarchical Transformer~\cite{masumura2021hierarchical} (H-Transformer), 
which consists of a text encoder with 4 transformer layers and a Conformer ASR module in our setups. 

\subsection{Results}
Table \ref{tab:all_results} shows the results of our experiments. We can find that both the role VAE and topic VAE show superior results on the final recognition accuracy. By combining them together, we can even achieve further improvement.  

Since the data set contains open-domain topics, the session-level language model makes the final recognition result worse on HKUST.
In addition, 
we also find that the topic-based rescoring operation has a positive effect on both data sets. 
Meanwhile, on the open-domain data set HKUST, the topic model rescoring is worse than that of the DDT data set with more obvious topic consistency. 

In general, we can find that after adding the variational module and the rescoring module, the recognition performance has been greatly improved, resulting a relative 12\% improvement on DDT set. 

\subsection{Context Length Analysis}
In a conversation, even on the same topic, 
as the number of conversation rounds increases, the speaker's speaking habits and the topics they are talking about will also change. 
At the same time, 
more recent texts may contain historical information that is more helpful for the recognition of the current sentence.
Therefore, we explore the role context length $j$ and topic context length $k$ of the input for the VAEs respectively on the HKUST dataset.

As shown in Table \ref{tab:role_length} and \ref{tab:topic_length}, 
we design experiments to explore the impact of context length on model performance, and find that when $k=3$ or $j=2$, 
the proposed model reaches the lowest CERs.
\begin{table}[htbp]
\vspace{-0.2cm}
\centering
\caption{The influence of role context length in the number of sentences on recognition accuracy with no topic context.}
\label{tab:role_length}
\begin{tabular}{lccc}
\hline
Role Context Length & 1     & 2     & 3     \\ \hline
\hline
CER(\%)        & 20.01 & 19.94 & 19.99 \\ \hline
\vspace{-1.2cm}
\end{tabular}
\end{table}
\begin{table}[htbp]
\centering
\caption{The influence of topic context length in the number of sentences on recognition accuracy with no role context.}
\label{tab:topic_length}
\begin{tabular}{lccc}
\hline
Topic Context Length & 1     & 2     & 3     \\ \hline
\hline
CER(\%)        & 20.05 & 19.96 & 19.81 \\ \hline
\vspace{-0.8cm}
\end{tabular}
\end{table}
\section{Conclusion}
This paper proposes a novel model to learn conversation-level characteristics including role preference and topical coherence in conversational speech recognition. 
We design a latent variational module (LVM) which contains two specific VAEs to learn the role preference and topical coherence. Meanwhile, a topic model is used on this basis to rescore the top-k outputs of the ASR model, biasing the results to words used in specific topics. 
Experimental results on conversational ASR tasks indicate that the proposed method effectively improves ASR performance.

\small
\bibliographystyle{IEEEbib}
\bibliography{strings,refs}

\end{document}